\documentclass[12pt]{article}
\usepackage{latexsym}
\begin{document}
\title{ \vspace{0.5cm} Hierarchy of massive gauge fields.}
\author{A.A.Slavnov.\\Steklov Mathematical Institute
\\Gubkina st. 8, 119991 Moscow,\& Moscow State University}
\maketitle

\begin{abstract}
An explicitely gauge invariant polynomial action for massive gauge fields
is proposed. For different values of parameters it describes massive Yang-Mills field, 
the Higgs-Kibble model, the model with spontaneously broken symmetry and two scalar mesons.
\end{abstract}

\section{Introduction}

Gauge invariance forbids a naive introduction of a mass term into the action of a vector field.
To preserve gauge invariance in the presence of a mass term it is necessary to increase the number of fields
entering the action functional.

This possibility is realized for example in the Stuekelberg formalism \cite{St}, using of which in non-Abelian gauge theories
is described in details in the paper \cite{Sl1}. In this formalism gauge invariance is achieved by changing 
the fields entering the action for massive vector fields by gauge transformed fields
\begin{equation}
A_{\mu} \rightarrow \Omega A_{\mu} \Omega^++ \partial_{\mu} \Omega \Omega^+
\label{1}
\end{equation}
where $\Omega(x)$ is an element of the gauge group, for example
$\Omega(x)=\exp \{iT^a \varphi^a(x) \}$, and $T^a$ are the generators of the corresponding algebra.
The group coordinates $ \varphi^a$ are considered as dynamical variables. If the kinetic part and
interaction are gauge invariant, all the dependence on the fields $\varphi$ is contained in the mass term
\begin{equation}
\frac{m^2}{2}(A_{\mu}+ \partial_{\mu} \Omega \Omega^+)^2
\label{3}
\end{equation}
In the unitary gauge $\varphi^a=0$ the dependence of the action on the additional fields disappears. 
The propagator of the massive vector field does not decrease at large momenta
leading to nonrenormalizability of the theory.
\begin{equation}
D_c^{\mu \nu}(k)= \frac{g^{\mu \nu}-k^{\mu}k^{\nu}m^{-2}}{k^2-m^2}
\label{4}
\end{equation}

Using the gauge invariance of the action one can impose another gauge condition, for example
 the Lorentz condition $\partial_{\mu}A_{\mu}=0$. It improves the ultraviolet asymptotics of the propagator
\begin{equation}
\tilde{D}_c^{\mu \nu}(k)= \frac{g^{\mu \nu}-k^{\mu}k^{\nu}k^{-2}}{k^2-m^2}
\label{5}
\end{equation}
however in this case the interaction Lagrangian includes vertices with arbitrary number of scalar
fields $\varphi^a$, which also does not allow to renormalize the theory.
The only exception is the theory with the gauge group $U(1)$, in which the auxilliary fields decouple
 completely, making the theory renormalizable.

The obvious disadvantage of the Stuekelberg formalism for non-Abelian theories is a nonpolynomial
structure of the action which includes an infinite series of vertices describing the interaction 
of the scalar field $\varphi$ (for recent discussion see \cite{FQ}).
 
These difficulties are absent in the model, where the vector field mass arises as a result
of spontaneous symmetry breaking by means of the Higgs mechanism \cite{Hi}, \cite{Ki}.

The corresponding Lagrangian includes the interaction with the scalar field and in the case when the 
complex scalar field $\varphi$ belongs to the fundamental representation of the gauge group $ SU(2)$
looks as follows
\begin{equation}
L=L_{YM}+(D_{\mu} \varphi)^+(D_{\mu} \varphi)- \lambda^2(\varphi^+ \varphi- \mu^2)^2
\label{6}
\end{equation}
where $L_{YM}$ denotes the usual Yang-Mills field Lagrangian, and the perturbation theory is developped near
the stable minimum
\begin{equation}
\varphi_1(x)=0, \varphi_2(x)= \mu \quad \mu= \mu^+
\label{7}
\end{equation}
In this case gauge invariance allows to eliminate three of four real components of the field $\varphi$.
The corresponding gauge is unitary as in this case the spectrum contains only physical exitations
  describing massive vector field and one real scalar field. As in the previous case the propagator of
 the gauge field does not decrease at large momenta leading to formal nonrenormalizability of the theory.

However, contrary to the model discussed above, in the Lorentz gauge the propagators of all fields
 decrease at infinity as $k^{-2}$,
and the interaction includes only the four point vertex without derivatives and the three point vertex
with one derivative. Such a model is renormalizable \cite{Ho}, and due to gauge invariance the
renormalized scattering matrix for a proper choice of counterterms coincides with the scattering matrix
in the unitary gauge \cite{Sl2}, \cite{T}, \cite{LZ}, \cite{HV}.

This mechanism is the basis of the Weinberg-Salam model of weak and electromagnetic interactions.
It is known that up to now all the attempts to find experimentally the scalar Higgs meson did not lead
 to success.

It is worth to mention that, as follows from the Lagrangian (\ref{6}) increasing of the Higgs meson mass
makes stronger the self interaction of the scalar fields and in the limit $m \rightarrow \infty$
the model coincides with the nonrenormalizable massive vector field in the Stuekelberg formalism.

The goal of this paper is to construct a gauge invariant polynomial action including only constraints
 related to gauge invariance, which for a
 different choice of parameters describes massive Yang-Mills field, Higgs-Kibble model, or its
 generalization, including additional scalar fields.

Such an action may be written in the form
\begin{eqnarray}
A= \int dx \{L_{YM}+(D_{\mu} \varphi)^+(D_{\mu} \varphi)+ \frac{1}{2\mu} 
\partial_{\mu}X \partial_{\mu}(\varphi^+ \varphi)+ \nonumber\\
\frac{a}{2} \Box{X} \Box{X}+ \frac{b}{2} \partial_{\mu} \Box{X} \partial_{\mu} \Box{X}
+ \partial_{\mu} \bar{c} \partial_{\mu}c \}
\label{9}
\end{eqnarray}
where for the fields $\varphi$ one has to use a standard decomposition
\begin{equation}
\varphi_1= \frac{iB_1+B_2}{\sqrt{2}}; \quad \varphi_2=\mu+ \frac{1}{ \sqrt{2}}( \sigma'-iB_3)
\label{10}
\end{equation}

The action (\ref{9}) includes a Hermitean scalar field $X$
and anticommuting ghost fields $c, \bar{c}$. It is obviously gauge invariant if
the fields $X$ and $c, \bar{c}$ are singlets of the gauge group.
Substituting into the action (\ref{9}) the decomposition
(\ref{10}) and introducing instead of the fields $X, \sigma$ their linear combinations
$X\equiv X, \sigma= \sigma'+X$ one can rewrite it in the form
\begin{eqnarray}
A= \int dx \{L_{YM}+ \frac{m^2}{2}A_{\mu}^2+mA_{\mu}^a \partial_{\mu}B^a+ 
\frac{1}{2} \partial_{\mu}B^a \partial_{\mu}B^a+ \nonumber\\
\frac{1}{2} \partial_{\mu} \sigma \partial_{\mu} \sigma- \frac{1}{2} \partial_{\mu}X \partial_{\mu} X
+\partial_{\mu} \bar{c} \partial_{\mu}c+ \frac{a}{2} \Box{X} \Box{X} 
+ \frac{b}{2} \partial_{\mu} \Box{X} \partial_{\mu} \Box{X}+ \nonumber\\
\frac{g}{2}A_{\mu}^a( \tilde{\sigma} \partial_{\mu}B^a-B^a \partial_{\mu} \tilde {\sigma}
- \epsilon^{abc}B^b \partial_{\mu}B^c)+ \nonumber\\
+ \frac{mg}{2} \tilde{\sigma}A_{\mu}^2+ \frac{g^2}{8}(\tilde{\sigma}^2+B^2)A_{\mu}^2+ 
\frac{g}{4m} \partial_{\mu}X \partial_\mu(B^2+ \tilde{\sigma}^2) \}
\label{11}
\end{eqnarray}
Here $ \tilde{\sigma}= \sigma-X$, and $m= \frac{\mu g}{\sqrt{2}}$

The action (\ref{11}) includes higher derivatives, however it is also invariant with
 respect to some supersymmetry transformation which leads to the conservation of a nilpotent
charge $Q$. Being restricted to the space of states satisfying the condition
\begin{equation}
Q| \Phi>=0
\label{12}
\end{equation}
the scattering matrix, corresponding to the action (\ref{11}) is unitary in the space
 including only nonnegative norm and energy states and for  $a=b=0$ it
provides a gauge invariant formulation of massive Yang-Mills field, for $b=0, a
\neq 0$ describes the Higgs-Kibble model, and for  $a \neq 0, b \neq 0$
it corresponds to the massive vector model with spontaneously broken symmetry and two scalar mesons.
We emphasize that for $a=b=0$ the action (\ref{11})
is polynomial and in distinction of the Stuekelberg formalism includes only finite number of
interaction  vertices. Moreover the transition from the Higgs model to the massive Yang-Mills
 field may be done at the level of Lagrangian by putting the parameters  $a$ and
$b$ equal to zero.

In the next sections we consider in details the model (\ref{11}) for
different values of the parameters $a$ and $b$.

\section{Massive Yang-Mills field.}
In this section we study the action (\ref{11}) for $a=b=0$.
This action may be written in the form
\begin{eqnarray}
A= \int dx \{\tilde{L}(A_{\mu}, B, \tilde{\sigma})+ \frac{1}{2} \partial_{\mu} \sigma \partial_{\mu} \sigma
- \frac{1}{2} \partial_{\mu}X \partial_{\mu}X+\nonumber\\
+ \partial_{\mu} \bar{c} \partial_{\mu}c + \frac{g}{4m} \partial_{\mu}X \partial_{\mu}(B^2+ \tilde{\sigma}^2)\}
\label{13}
\end{eqnarray}
where $\tilde{L}+1/2 \partial_{\mu} \sigma \partial_{\mu} \sigma$ is the usual
 Lagrangian for Yang-Mills field interacting with $B$ and $ \sigma$.

This action may be obtained from the Stuekelberg-like action, including the constraint $ \varphi^+ \varphi= \mu^2$
\begin{equation}
A_{St}= \int dx \{L_{YM}+(D_{\mu} \varphi)^+(D_{\mu} \varphi)+ \frac{1}{2 \mu}X(\varphi^+ \varphi- \mu^2)
\label{14}
\end{equation}
by changing variables
\begin{equation}
X \rightarrow \Box{X}
\label{15}
\end{equation}
and shifting the field $\sigma$, $ \sigma \rightarrow \sigma+X+ \sqrt{2}\mu$. 
It was shown in our paper \cite{Sl3}, that to preserve the equivalence of the original theory and
the theory obtained after the change $\phi \rightarrow \phi(\phi')$
it is necessary to modify the original Lagrangian, adding the interaction with the ghost fields $\bar{c}, c$
\begin{equation}
A(\varphi) \rightarrow \tilde{A}(\varphi', \bar{c},c)= \int dx \{ \bar{c} \frac{\delta \varphi}{\delta \varphi'}c+L[\varphi(\varphi')] \}
\label{16}
\end{equation}
The action $\tilde{A}(\phi, \bar{c},c)$ is invariant with respect to the supersymmetry transformations
\begin{eqnarray}
\delta \varphi'=c \epsilon; \quad \delta c=0 \nonumber\\
\delta \bar{c}= \frac{\delta A}{ \delta \varphi(x)}( \varphi') \epsilon
\label{17}
\end{eqnarray}
This invariance leads to existence of the conserved nilpotent charge $Q$, 
annhilating the invariant
subspace (\ref{12}) with nonnegative metrics. For the model under consideration this transformation has a form
\begin{eqnarray}
\delta X=c \epsilon; \quad 
\delta \sigma=c \epsilon \nonumber\\
\delta \bar{c}=X \epsilon- \sigma \epsilon- \frac{g}{4m}(B^2+ \tilde{\sigma}^2) \epsilon
\label{18}
\end{eqnarray}

In the unitary gauge $B^a=0$ the action (\ref{13}) describes massive particles with spin equal
 to one, scalar field with the positive norm $(\sigma)$ and the scalar field with the negative norm $(X)$.
For asymptotic states the condition (\ref{12}) linearizes and the charge $Q$ acquires a form
\begin{equation}
Q_0= \int d^3k \{c^+(k)(\sigma^-(k)-X^-(k))+(\sigma^+(k)-X^+(k))c^-(k) \}
\label{19}
\end{equation}
where $c^{\pm}, \sigma^{\pm}, X^{\pm}$ are the creation and annihilation operators for the fields $c, \sigma, X$
satisfying the commutation relations
\begin{eqnarray}
[\sigma^-(k), \sigma^+(k')]= \delta(k-k'); \quad [X^-(k),X^+(k')]=- \delta(k-k') \nonumber\\
\{c^-(k), \bar{c}^+(k') \}= \delta (k-k'); \quad \{\bar{c}^-(k), c^+(k') \}= \delta(k-k')
\label{20}
\end{eqnarray}
In our case the fields $c, \bar{c}$ are free, so the condition
 (\ref{12}) is reduced to the following equation
\begin{equation}
( \sigma^-(k)-X^-(k))| \Phi>=0
\label{21}
\end{equation}
The solution of this equation is
 \begin{equation}
| \phi>=[1+ \Sigma_n c_n \Pi_{1 \leq j \leq n}( \sigma^+(k_j)-X^+(k_j))]|\Phi>_A
\label{22}
\end{equation}
where the vector $|\Phi>_A$ describes the states of the massive vector field. The vectors (\ref{22}) obviously
have nonnegative norm and the matrix elements of any operator commuting with $Q$,
calculated in the space of the vectors $|\Phi>$ , coincide with the corresponding matrix elements
in the space $|\Phi>_A$. It proves the unitarity of the model in the physical subspace.

Gauge invariance and polynomial structure of the action(\ref{13}) is obvious.
We stress that the polynomial structure of the action (\ref{13}) does not guarantee renormalizability,
 as the interaction vertices include derivatives. One may hope however that the existence of the
 explicitely gauge invariant polynomial action will simplify the analysis of this model.

\section{Higgs model}

In this section we consider the action (\ref{9}) for $b=0, a \neq 0$.
Being written in the variables $B, \sigma$ this action differs from (\ref{13})
by the additional term with higher derivatives:
\begin{eqnarray}
A= \int dx \{\tilde{L}(A_{\mu},B, \tilde{\sigma})+ \frac{1}{2} \partial_{\mu}
 \sigma \partial_{\mu} \sigma- \frac{1}{2} \partial_{\mu}X \partial_{\mu}X +\nonumber\\
\frac{a}{2} \Box{X} \Box{X}+ \partial_{\mu} \bar{c} \partial_{\mu}c+ \frac{g}{4m} 
\partial_{\mu}X \partial_{\mu}(B^2+ \tilde{\sigma}^2)
\label{23}
\end{eqnarray}

To quantize this action one has to use the Ostrogradsky canonical formulation for theories
with higher derivatives \cite{Sl4}.

We choose again the unitary gauge $B=0$.
To develop the canonical formulation we introduce the variables
$Q_1=X, \quad Q_2= \dot{X}$ and conjugated momenta
\begin{equation}
P_1=- \partial_0X-a \partial_0 \Box{X}+ \tilde{f}(A_{\mu}, \sigma,X); \quad
P_2=a \Box{X}
\label{24}
\end{equation}
Here $ \tilde{f}$ denotes the terms of order $g$.

In terms of these variables the Fourier transform of the Hamiltonian looks as follows
\begin{eqnarray}
H=P_1Q_2+P_2 \dot{Q}_2-L= \nonumber\\
P_1Q_2-k^2P_2Q_1+ \frac{P_2^2}{2a}+ \frac{1}{2}Q_2^2- \frac{1}{2}k^2Q_1^2+ 
\frac{P_{\sigma}^2}{2}+ \nonumber\\
\frac{1}{2}k^2 \sigma^2+p_cp_{\bar{c}}+k^2 \bar{c}c+ \tilde{H}(A_{\mu}, 
\tilde{\sigma},X)
\label{25}
\end{eqnarray}
In this equation $ \tilde{H}$ denotes the Hamiltonian of the free massive vector field and the 
interaction terms. For the fields $\sigma$ and $c, \bar{c}$
we use standard canonical variables.

The spectrum of the free Hamiltonian includes two bosonic states with zero masses,
 one bosonic state with the mass $a^{-1/2}$ and
massless states corresponding to the ghost fields.

The corresponding creation and annihilation operators look as follows
\begin{equation}
m=0: \quad a^{\pm}(k)= \frac{1}{\sqrt{2 \omega_1}}(P_1 \mp i \omega_1 Q_1 \mp
i \omega_1 P_2); \quad \omega_1= \sqrt{k^2} 
\label{26}
\end{equation}
\begin{equation}
[a^-(k),a^+(k')]=- \delta(k-k')
\label{26a}
\end{equation}
\begin{equation}
m^2=a^{-1}: \quad b^{ \pm}= \frac{1}{\sqrt{2 \omega_2}}(P_1+Q_2 \mp i \omega_2P_2)
\quad \omega_2= \sqrt{k^2+a^{-1}}
\label{27a}
\end{equation}
\begin{equation}
[b^-(k),b^+(k')]= \delta(k-k')
\label{27}
\end{equation}
\begin{equation}
m=0: \quad \sigma^{ \pm}= \frac{1}{\sqrt{2 \omega_1}}( \omega_1 \sigma \mp ip_{\sigma}) \quad 
[\sigma^-, \sigma^+]= \delta(k-k')
\label{28}
\end{equation}

One sees that the operators $a^+$ create the massless states with the negative norm.
However, as in the case of the massive vector field, due to invariance of the action (\ref{23})
with respect to the supersymmetry transformation
\begin{eqnarray}
\delta X=c \epsilon; \quad 
\delta \sigma=c \epsilon \nonumber\\
\delta \bar{c}=X \epsilon- \sigma \epsilon-a \Box{X}-\frac{g}{4m}(B^2+
\tilde{\sigma}^2)
\label{29}
\end{eqnarray}
there exists a conserved nilpotent operator $Q$, and the invariant subspace 
 (\ref{12}) has a nonnegative norm.

The asymptotic operator $Q_0$ has a form
\begin{eqnarray}
Q_0= \int d^3k(Q_1 \partial_0 c+P_1c+P_2 \partial_0 c-p_{\sigma}c+ 
\sigma \partial_0 c)=\nonumber\\
\int d^3k [c^+(k)(a^-(k)- \sigma^-(k))+(a^+(k)- \sigma^+(k))c^-(k)]
\label{30}
\end{eqnarray}
As in the previous case the condition (\ref{12}) reduces to the equation
\begin{equation}
(a^-- \sigma^-)| \Phi>=0
\label{31}
\end{equation}
The solution of this equation looks as follows:
\begin{equation}
|\Phi>=[1+ \Sigma_n c_n \Pi_{1 \leq j \leq n}( \sigma^+(k_j)-a^+(k_j))]|\Phi>_{A,b}
\label{32}
\end{equation}
where the vectors $|\Phi>_{A,b}$ describe the states of the massive vector field and the scalar particle
 with the mass  $a^{-1/2}$.

It is easy to see that in the Lorentz gauge $\partial_{\mu}A_{\mu}=0$ due to the presence of higher
 derivatives of the field $X$ the model is renormalizable and coincides with the Higgs-Kibble model.

\section{The model with spontaneously broken symmetry and two scalar mesons.}
As the last example we consider the action (\ref{11})
for $a,b \neq 0$. We use again the unitary gauge $B=0$.
The action takes the form
\begin{eqnarray}
A= \int dx \{\tilde{L}(A_{\mu}, \tilde{\sigma})+ 
\frac{1}{2} \partial_{\mu} \sigma \partial_{\mu} \sigma- \frac{1}{2} \partial_{\mu}X \partial_{\mu}X
+\frac{a}{2} \Box{X} \Box{X}+\nonumber\\
\frac{b}{2} \partial_{\mu} \Box{X} \partial_{\mu} \Box{X}+ \partial_{\mu} \bar{c} \partial_{\mu}c+
\frac{g}{4m} \partial_{\mu}X \partial_{\mu}(\tilde{\sigma}^2)
\label{33}
\end{eqnarray}

To construct the canonical formalism we introduce the variables
\begin{equation}
Q_1=X; \quad Q_2= \partial_0X; \quad Q_3= \partial_0^2X
\label{34}
\end{equation}
and the conjugated momenta
\begin{eqnarray}
P_3=b \partial_0 \Box{X}; \quad 
P_2=a \Box{X}-b \partial_0 \Box{X} \nonumber\\
P_1=- \partial_0 X-b \partial_i^2 \partial_0 \Box{X}-a \partial_0 \Box{X}+ \partial_0^3 \Box{X}
+ \tilde{f}(A_{\mu},X, \sigma)
\label{35}
\end{eqnarray}
The Hamiltonian looks as follows
\begin{eqnarray}
H=P_1Q_2+P_2Q_3+P_3 \dot{Q}_3-L= \nonumber\\
P_1Q_2+P_2Q_3-k^2P_3Q_2+ \frac{P_3^2}{2b}+ \frac{b}{2}(Q_3+k^2Q_1)k^2(Q_3+k^2Q_1)- \nonumber\\
\frac{a}{2}(Q_3+k^2Q_1)^2+ \frac{1}{2}Q_2^2- \frac{1}{2}k^2Q_1^2+ \frac{p_{\sigma}^2}{2}+ 
\frac{1}{2}k^2 \sigma^2+p_cp_{\bar{c}}+k^2\bar{c}c + \tilde{H}
\label{36}
\end{eqnarray}
where $ \tilde{H}$  denotes the Hamiltonian of the free vector field and the interaction terms.
The spectrum of the free Hamiltonian includes 
one massless positive norm state, one massless negative norm state and two exitations
 with the masses
\begin{equation}
M^2= \frac{-a \pm \sqrt(a^2+4b)}{2b}
\label{37}
\end{equation}
For $a>0, b<0$ squares of both masses and the norms of both states are positive.
The theory is invariant with respect to the supersymmetry transformations
\begin{eqnarray}
\delta X=c \epsilon; \quad 
\delta \sigma=c \epsilon \nonumber\\
\delta \bar{c}=X \epsilon- \sigma \epsilon-a \Box{X}+b (\Box^2){X}- \frac{g}{4m}(B^2+ \tilde{\sigma}^2)
\label{38}
\end{eqnarray}
The conserved charge $Q$ asymptotically coincides with the charge $Q_0$ in the Higgs model.
The physical state vectors look as in the model discussed above
\begin{equation}
|\Phi>=[1+ \Sigma_n c_n \Pi_{1 \leq j \leq n}(a^+(k_j)- \sigma^+(k_j))]|\Phi>_f
\label{40}
\end{equation}
where $|\Phi>_f$ are the state vectors corresponding to the massive vector field
 and two scalar fields with the masses given by the eq.(\ref{37}).

As in the previous case in the gauge $B=0$ the theory is manifestly unitary.
Gauge invariance of the model allows to pass to other gauges, like the Lorentz gauge,
which makes the renormalizability obvious.

\section{Conclusion.}
In this paper we showed that the most interesting gauge invariant models of massive vector fields
may be described by the unique polynomial action, which includes only constraints related to
gauge invariance. The different values
of parameters entering this action correspond to the different models, namely
the massive vector field, the Higgs-Kibble model or the model with spontaneously
 broken symmetry and two scalar mesons.
As in the usual formalism renormalizable Higgs model for infinite value of the scalar
 meson mass transforms into nonrenormalizable theory of massive vector field.

Qualitatively an analogous behaviour is demonstrated by the model with two
scalar mesons. If the masses of both mesons tend to infinity, the model transforms into the
nonrenormalizable massive vector field theory. However the limitations on the masses of the scalar
mesons in this case differ from the Higgs model, which may be of some interest
from the point of view of experiment.

The same formalism may be used to write the polynomial Lagrangian for the 
nonlinear sigma-model. 
It is given by the eq.(\ref{13}), with $A_{\mu}=0$. 

{\bf Acknowledgements} \\
This work was completed while the author was visiting University of Milan.
I thank R.Ferrari for hospitality and Cariplo foundation
 for support in the framework of Landau-Volta network. Useful discussions 
with R.Ferrari and A.Quadri are acknowledged.
This work was partially supported by RBRF under grant 050100541,
grant for support of leading scientific schools 20052.2003.1,
and the program "Theoretical problems of mathematics".
 \end{document}